% hep-th/0411271, v1
%

\input harvmac

\noblackbox

\def\IZ{\relax\ifmmode\mathchoice
{\hbox{\cmss Z\kern-.4em Z}}{\hbox{\cmss Z\kern-.4em Z}}
{\lower.9pt\hbox{\cmsss Z\kern-.4em Z}}
{\lower1.2pt\hbox{\cmsss Z\kern-.4em Z}}\else{\cmss Z\kern-.4em Z}\fi}
\def\IB{\relax{\rm I\kern-.18em B}}
\def\IC{{\relax\hbox{\kern.3em{\cmss I}$\kern-.4em{\rm C}$}}}
\def\ID{\relax{\rm I\kern-.18em D}}
\def\IE{\relax{\rm I\kern-.18em E}}
\def\IF{\relax{\rm I\kern-.18em F}}
\def\IG{\relax\hbox{$\inbar\kern-.3em{\rm G}$}}
\def\IGa{\relax\hbox{${\rm I}\kern-.18em\Gamma$}}
\def\IH{\relax{\rm I\kern-.18em H}}
\def\II{\relax{\rm I\kern-.18em I}}
\def\IK{\relax{\rm I\kern-.18em K}}
\def\IP{\relax{\rm I\kern-.18em P}}
%\def\IX{\relax{\rm X\kern-.01em X}}
%this doesn't work

\font\cmss=cmss10 \font\cmsss=cmss10 at 7pt
\def\IR{\relax{\rm I\kern-.18em R}}

\def\frac#1#2{{#1 \over #2}}

\def\OL#1{ \kern1pt\overline{\kern-1pt#1
     \kern-1pt}\kern1pt }

\def\tb{\overline{\tau}}

%\GopakumarMU
\lref\GopakumarMU{ R.~Gopakumar and S.~Mukhi, ``Orbifold and orientifold compactifications of F-theory and
M-theory  to six and four dimensions,'' Nucl.\ Phys.\ B {\bf 479}, 260 (1996) [arXiv:hep-th/9607057].
%%CITATION = HEP-TH 9607057;%%
}

\lref\DDFK{F. Denef, M. Douglas, B. Florea, and S. Kachru, in progress.}

%\BlumenhagenVR
\lref\BlumenhagenVR{ R.~Blumenhagen, D.~Lust and T.~R.~Taylor, ``Moduli stabilization in chiral type IIB
orientifold models with fluxes,'' Nucl.\ Phys.\ B {\bf 663}, 319 (2003) [arXiv:hep-th/0303016].
%%CITATION = HEP-TH 0303016;%%
}

\lref\ohta{
%\OhtaWK
N.~Ohta, ``Accelerating Cosmologies and Inflation from M/Superstring Theories,'' arXiv:hep-th/0411230.
%%CITATION = HEP-TH 0411230;%%
}

\lref\gia{
%\DvaliTM
G.~Dvali, ``Large hierarchies from attractor vacua,'' arXiv:hep-th/0410286.
%%CITATION = HEP-TH 0410286;%%
}

%\CamaraJJ
\lref\CamaraJJ{ P.~G.~Camara, L.~E.~Ibanez and A.~M.~Uranga, ``Flux-induced SUSY-breaking soft terms on D7-D3
brane systems,'' arXiv:hep-th/0408036;
%%CITATION = HEP-TH 0408036;%%
%\LustFI
D.~Lust, S.~Reffert and S.~Stieberger, ``Flux-induced soft supersymmetry breaking in chiral type IIb
orientifolds with D3/D7-branes,'' arXiv:hep-th/0406092.
%%CITATION = HEP-TH 0406092;%%
}

\lref\beckersX{
%\BeckerGJ
K.~Becker and M.~Becker, ``M-Theory on Eight-Manifolds,'' Nucl.\ Phys.\ B {\bf 477}, 155 (1996)
[arXiv:hep-th/9605053].
%%CITATION = HEP-TH 9605053;%%
}

\lref\distributions{S.~Ashok and M.~R.~Douglas, ``Counting flux vacua,'' JHEP {\bf 0401}, 060 (2004)
[arXiv:hep-th/0307049]. \
%%CITATION = HEP-TH 0307049;%%
F.~Denef and M.~R.~Douglas, ``Distributions of flux vacua,'' JHEP {\bf 0405}, 072 (2004) [arXiv:hep-th/0404116].
%%CITATION = HEP-TH 0404116;%%
A.~Giryavets, S.~Kachru and P.~K.~Tripathy, ``On the taxonomy of flux vacua,'' arXiv:hep-th/0404243; O. de
Wolfe, A. Giryavets, S. Kachru, W. Taylor, ``Enumerating enhanced symmetries in flux vacua",
arXiv:hep-th/0411061. }

%%CITATION = HEP-TH 0411061;%%
%%CITATION = HEP-TH 0404243;%%

\lref\susyscale{
%\BanksES
T.~Banks, M.~Dine and E.~Gorbatov, ``Is there a string theory landscape?,'' arXiv:hep-th/0309170;
%%CITATION = HEP-TH 0309170;%%
%\DineIS
M.~Dine, E.~Gorbatov and S.~Thomas,
%``Low energy supersymmetry from the landscape,''
arXiv:hep-th/0407043;
%%CITATION = HEP-TH 0407043;%%
%\DouglasQG
M.~R.~Douglas, ``Statistical analysis of the supersymmetry breaking scale,'' arXiv:hep-th/0405279.
%%CITATION = HEP-TH 0405279;%%
%\SusskindUV
L.~Susskind, ``Supersymmetry breaking in the anthropic landscape,'' arXiv:hep-th/0405189;
%%CITATION = HEP-TH 0405189;%%
E. Silverstein, ``Counter-intuition and Scalar Masses", arXiv:hep-th/0407202.
%%CITATION = HEP-TH 0407202;%%
 }

%\BeckerGW
\lref\BeckerGW{ M.~Becker, G.~Curio and A.~Krause, ``De Sitter vacua from heterotic M-theory,'' Nucl.\ Phys.\ B
{\bf 693}, 223 (2004) [arXiv:hep-th/0403027].
%%CITATION = HEP-TH 0403027;%%
}

\lref\us{
%\SaltmanSN
A.~Saltman and E.~Silverstein, ``The scaling of the no-scale potential and de Sitter model building,''
arXiv:hep-th/0402135.
%%CITATION = HEP-TH 0402135;%%
 }

%\MarchesanoYN
\lref\MarchesanoYN{ F.~Marchesano, G.~Shiu and L.~T.~Wang, ``Model building and phenomenology of flux-induced
supersymmetry breaking on D3-branes,'' arXiv:hep-th/0411080.
%%CITATION = HEP-TH 0411080;%%
}

\lref\kreuzer{ http://hep.itp.tuwien.ac.at/~kreuzer/CY/ }

\lref\beckers{
%\BeckerNN
K.~Becker, M.~Becker, M.~Haack and J.~Louis, ``Supersymmetry breaking and alpha'-corrections to flux induced
potentials,'' JHEP {\bf 0206}, 060 (2002) [arXiv:hep-th/0204254].
%%CITATION = HEP-TH 0204254;%%
 }

%\KlemmTS
\lref\KlemmTS{ A.~Klemm, B.~Lian, S.~S.~Roan and S.~T.~Yau, ``Calabi-Yau fourfolds for M- and F-theory
compactifications,'' Nucl.\ Phys.\ B {\bf 518}, 515 (1998) [arXiv:hep-th/9701023].
%%CITATION = HEP-TH 9701023;%%
}

\lref\svw{
%\SethiES
S.~Sethi, C.~Vafa and E.~Witten, ``Constraints on low-dimensional string compactifications,'' Nucl.\ Phys.\ B
{\bf 480}, 213 (1996) [arXiv:hep-th/9606122].
%%CITATION = HEP-TH 9606122;%%
}

\lref\inflow{
%\CheungAZ
Y.~K.~Cheung and Z.~Yin, ``Anomalies, branes, and currents,'' Nucl.\ Phys.\ B {\bf 517}, 69 (1998)
[arXiv:hep-th/9710206].
%%CITATION = HEP-TH 9710206;%%
%\MinasianMM
R.~Minasian and G.~W.~Moore, ``K-theory and Ramond-Ramond charge,'' JHEP {\bf 9711}, 002 (1997)
[arXiv:hep-th/9710230].
%%CITATION = HEP-TH 9710230;%%
 }

\lref\splitsusy{%\ArkaniHamedFB
N.~Arkani-Hamed and S.~Dimopoulos, ``Supersymmetric unification without low energy supersymmetry and signatures
for fine-tuning at the LHC,'' arXiv:hep-th/0405159;
%%CITATION = HEP-TH 0405159;%%
%\GiudiceTC
G.~F.~Giudice and A.~Romanino, ``Split supersymmetry,'' Nucl.\ Phys.\ B {\bf 699}, 65 (2004)
[arXiv:hep-ph/0406088].
%%CITATION = HEP-PH 0406088;%%
}

\lref\luty{%\LutyFF
M.~A.~Luty, ``Weak scale supersymmetry without weak scale supergravity,'' Phys.\ Rev.\ Lett.\  {\bf 89}, 141801
(2002) [arXiv:hep-th/0205077].
%%CITATION = HEP-TH 0205077;%%
}

\lref\fluxorientifold{%\AngelantonjRQ
C.~Angelantonj, S.~Ferrara and M.~Trigiante, ``New D = 4 gauged supergravities from N = 4 orientifolds with
fluxes,'' JHEP {\bf 0310}, 015 (2003) [arXiv:hep-th/0306185].
%%CITATION = HEP-TH 0306185;%%
}

\lref\AR{%\KalloshYH
R.~Kallosh and A.~Linde, ``Landscape, the scale of SUSY breaking, and inflation,'' arXiv:hep-th/0411011.
%%CITATION = HEP-TH 0411011;%%
}

\lref\keshavetal{%\DasguptaSS
K.~Dasgupta, G.~Rajesh and S.~Sethi, ``M theory, orientifolds and G-flux,'' JHEP {\bf 9908}, 023 (1999)
[arXiv:hep-th/9908088].
%%CITATION = HEP-TH 9908088;%%
}

\lref\senprobe{%\SenGV
A.~Sen, ``Orientifold limit of F-theory vacua,'' Phys.\ Rev.\ D {\bf 55}, 7345 (1997) [arXiv:hep-th/9702165].
%%CITATION = HEP-TH 9702165;%%
}

\lref\dealwisetal{
%\MyersFV
R.~C.~Myers, ``New Dimensions For Old Strings,'' Phys.\ Lett.\ B {\bf 199}, 371 (1987).
%%CITATION = PHLTA,B199,371;%%
%\deAlwisPR
S.~P.~de Alwis, J.~Polchinski and R.~Schimmrigk, ``Heterotic Strings With Tree Level Cosmological Constant,''
Phys.\ Lett.\ B {\bf 218}, 449 (1989).
%%CITATION = PHLTA,B218,449;%%
 }

\lref\fpapers{%\VafaXN
C.~Vafa, ``Evidence for F-Theory,'' Nucl.\ Phys.\ B {\bf 469}, 403 (1996) [arXiv:hep-th/9602022].
%%CITATION = HEP-TH 9602022;%%
;
%\MorrisonNA
D.~R.~Morrison and C.~Vafa, ``Compactifications of F-Theory on Calabi--Yau Threefolds -- I,'' Nucl.\ Phys.\ B
{\bf 473}, 74 (1996) [arXiv:hep-th/9602114].
%%CITATION = HEP-TH 9602114;%%
%\MorrisonPP
D.~R.~Morrison and C.~Vafa, ``Compactifications of F-Theory on Calabi--Yau Threefolds -- II,'' Nucl.\ Phys.\ B
{\bf 476}, 437 (1996) [arXiv:hep-th/9603161].
%%CITATION = HEP-TH 9603161;%%
%\BershadskyNH
M.~Bershadsky, K.~A.~Intriligator, S.~Kachru, D.~R.~Morrison, V.~Sadov and C.~Vafa, ``Geometric singularities and
enhanced gauge symmetries,'' Nucl.\ Phys.\ B {\bf 481}, 215 (1996) [arXiv:hep-th/9605200]
%%CITATION = HEP-TH 9605200;%%
}

%\SilversteinID
\lref\modlectures{ E.~Silverstein, ``TASI / PiTP / ISS lectures on moduli and microphysics,''
arXiv:hep-th/0405068; to appear in the proceedings of TASI 2003.
%%CITATION = HEP-TH 0405068;%%
}

\lref\senF{
%\SenBP
A.~Sen, ``Orientifold limit of F-theory vacua,'' Nucl.\ Phys.\ Proc.\ Suppl.\  {\bf 68}, 92 (1998) [Nucl.\ Phys.\
Proc.\ Suppl.\  {\bf 67}, 81 (1998)] [arXiv:hep-th/9709159].
%%CITATION = HEP-TH 9709159;%%
}

\lref\coulombduals{
%\FabingerGP
M.~Fabinger and E.~Silverstein, ``D-Sitter space: Causal structure, thermodynamics, and entropy,''
arXiv:hep-th/0304220;
%%CITATION = HEP-TH 0304220;%%
%\KarchEM
A.~Karch, ``Auto-localization in de-Sitter space,'' JHEP {\bf 0307}, 050 (2003) [arXiv:hep-th/0305192];
%%CITATION = HEP-TH 0305192;%%
%\SilversteinJP
E.~Silverstein, ``AdS and dS entropy from string junctions,'' arXiv:hep-th/0308175;
%%CITATION = HEP-TH 0308175;%%
%\AlishahihaMD
M.~Alishahiha, A.~Karch, E.~Silverstein and D.~Tong, ``The dS/dS correspondence,'' arXiv:hep-th/0407125.
%%CITATION = HEP-TH 0407125;%%
}

%\AndrianopoliSA
\lref\ferrara{ L.~Andrianopoli, S.~Ferrara and M.~Trigiante, ``Fluxes,
supersymmetry breaking and gauged
supergravity,'' arXiv:hep-th/0307139.
%%CITATION = HEP-TH 0307139;%%
}

%\MayrHH
\lref\MayrHH{ P.~Mayr,
  ``On supersymmetry breaking in string theory and its realization in brane
worlds,'' Nucl.\ Phys.\ B {\bf 593}, 99 (2001)
[arXiv:hep-th/0003198].
%%CITATION = HEP-TH 0003198;%%
}

%\TaylorII
\lref\TaylorII{ T.~R.~Taylor and C.~Vafa, ``RR flux on Calabi-Yau and
partial supersymmetry breaking,'' Phys.\
Lett.\ B {\bf 474}, 130 (2000) [arXiv:hep-th/9912152].
%%CITATION = HEP-TH 9912152;%%
}

%\GiryavetsVD
\lref\GiryavetsVD{ A.~Giryavets, S.~Kachru, P.~K.~Tripathy and
S.~P.~Trivedi, ``Flux compactifications on
Calabi-Yau threefolds,'' arXiv:hep-th/0312104.
%%CITATION = HEP-TH 0312104;%%
}

%\TripathyQW
\lref\TripathyQW{
P.~K.~Tripathy and S.~P.~Trivedi,
``Compactification with flux on K3 and tori,'' JHEP {\bf 0303}, 028 (2003)
[arXiv:hep-th/0301139].
%%CITATION = HEP-TH 0301139;%%
}

\lref\GVW{S. Gukov, C. Vafa and E. Witten, ``CFTs from Calabi-Yau
Fourfolds,'' Nucl. Phys. {\bf B584}, 69
(2000)[arXiv:hep-th/9906070]}

\lref\gkp{ S.~B.~Giddings, S.~Kachru and J.~Polchinski,` `Hierarchies from fluxes in string compactifications,''
Phys.\ Rev.\ D {\bf 66}, 106006 (2002) [arXiv:hep-th/0105097].
%%CITATION = HEP-TH 0105097;%%
}

\lref\kklt{ S.~Kachru, R.~Kallosh, A.~Linde and S.~P.~Trivedi,``De Sitter vacua in string theory,'' Phys.\ Rev.\ D
{\bf 68},046005 (2003) [arXiv:hep-th/0301240].
%%CITATION = HEP-TH 0301240;%%
}

\lref\KST{ S.~Kachru, M.~B.~Schulz and S.~Trivedi, ``Moduli stabilization
from fluxes in a simple IIB
orientifold,'' JHEP {\bf0310}, 007 (2003) [arXiv:hep-th/0201028].
%%CITATION = HEP-TH 0201028;%%
}

%\KachruNS
\lref\KachruNS{ S.~Kachru, X.~Liu, M.~B.~Schulz and S.~P.~Trivedi,
``Supersymmetry changing bubbles in string
theory,'' JHEP {\bf 0305}, 014 (2003) [arXiv:hep-th/0205108].
%%CITATION = HEP-TH 0205108;%%
}

%\FreyHF
\lref\FreyHF{ A.~R.~Frey and J.~Polchinski, ``N = 3 warped
compactifications,'' Phys.\ Rev.\ D {\bf 65}, 126009
(2002) [arXiv:hep-th/0201029].
%%CITATION = HEP-TH 0201029;%%
}

\lref\WSuper{E.~Witten,``Non-Perturbative Superpotentials In String
Theory,'' Nucl.\ Phys.\ B {\bf 474}, 343
(1996) [arXiv:hep-th/9604030].
%%CITATION = HEP-TH 9604030;%%
}

\lref\Candelas{ P.~Candelas and X.~de la Ossa,``Moduli Space Of Calabi-Yau
Manifolds,'' Nucl.\ Phys.\ B {\bf
355}, 455 (1991).
%%CITATION = NUPHA,B355,455;%%
}

\lref\mathematica{ S. Wolfram, ``The Mathematic Book,'' {\it Wolfram
Media/Cambridge Univ. Pr., 1999} }

\lref\JoeTwo{ J.~Polchinski, ``String Theory. Vol. 2: Superstring Theory And
Beyond,'' {\it  Cambridge, UK:
Univ. Pr. (1998) 531 p}.}

 % five point bold
      % seven point bold
                  % ten point bold
      % five point math bold
      % seven point math bold
                 % ten point math bold

\lref\mss{
%\MaloneyRR
A.~Maloney, E.~Silverstein and A.~Strominger, ``De Sitter space in
noncritical string theory,''
arXiv:hep-th/0205316;
%%CITATION = HEP-TH 0205316;%%
%\SilversteinXN
E.~Silverstein, ``(A)dS backgrounds from asymmetric orientifolds,''
arXiv:hep-th/0106209.
%%CITATION = HEP-TH 0106209;%%
}

\lref\trap{L. Kofman, A. Linde, X. Liu, A. Maloney, L. McAllister, and E.
Silverstein, ``Moduli Trapping from
Particle Production", to appear.}

\lref\dine{M. Dine, ``Towards a Solution of the Moduli Problems of String
Cosmology,'' Phys.Lett. {\bf{B}}482
(2000) 213, hep-th/0002047 \semi M. Dine, Y. Nir, and Y. Shadmi, ``Enhanced
Symmetries and the Ground State of
String Theory,'' Phys.Lett. {\bf{B}}438 (1998) 61, hep-th/9806124.}

%\BurgessIC
\lref\renata{ C.~P.~Burgess, R.~Kallosh and F.~Quevedo, ``de Sitter string
vacua from supersymmetric D-terms,''
JHEP {\bf 0310}, 056 (2003) [arXiv:hep-th/0309187].
%%CITATION = HEP-TH 0309187;%%
}

%\SilversteinJP
\lref\SilversteinJP{ E.~Silverstein, ``AdS and dS entropy from string
junctions,'' arXiv:hep-th/0308175.
%%CITATION = HEP-TH 0308175;%%
}

%\FabingerGP
\lref\FabingerGP{ M.~Fabinger and E.~Silverstein, ``D-Sitter
space: Causal structure, thermodynamics, and entropy,''
arXiv:hep-th/0304220.
%%CITATION = HEP-TH 0304220;%%
}

%\SusskindKW
\lref\SusskindKW{ L.~Susskind, ``The anthropic landscape of string theory,''
arXiv:hep-th/0302219.
%%CITATION = HEP-TH 0302219;%%
}

%\AshokGK
\lref\AshokGK{ S.~Ashok and M.~R.~Douglas, ``Counting flux vacua,''
arXiv:hep-th/0307049.
%%CITATION = HEP-TH 0307049;%%
}

\lref\achar{
%\AcharyaKV
B.~S.~Acharya, ``A moduli fixing mechanism in M theory,''
arXiv:hep-th/0212294.
%%CITATION = HEP-TH 0212294;%%
}

\lref\W{
%\WittenQJ
E.~Witten, ``Anti-de Sitter space and holography,'' Adv.\ Theor.\ Math.\
Phys.\  {\bf 2}, 253 (1998)
[arXiv:hep-th/9802150].
%%CITATION = HEP-TH 9802150;%%
}

%\StromingerSH
\lref\StromingerSH{ A.~Strominger and C.~Vafa, ``Microscopic Origin of the
Bekenstein-Hawking Entropy,'' Phys.\
Lett.\ B {\bf 379}, 99 (1996) [arXiv:hep-th/9601029].
%%CITATION = HEP-TH 9601029;%%
}

\lref\GKPads{
%\GubserBC
S.~S.~Gubser, I.~R.~Klebanov and A.~M.~Polyakov, ``Gauge theory correlators
from non-critical string theory,''
Phys.\ Lett.\ B {\bf 428}, 105 (1998) [arXiv:hep-th/9802109].
%%CITATION = HEP-TH 9802109;%%
}

%\KachruNS
\lref\KachruNS{ S.~Kachru, X.~Liu, M.~B.~Schulz and S.~P.~Trivedi,
``Supersymmetry changing bubbles in string
theory,'' JHEP {\bf 0305}, 014 (2003) [arXiv:hep-th/0205108].
%%CITATION = HEP-TH 0205108;%%
}

\lref\GH{
%\GibbonsMU
G.~W.~Gibbons and S.~W.~Hawking, ``Cosmological Event Horizons,
Thermodynamics, And Particle Creation,'' Phys.\
Rev.\ D {\bf 15}, 2738 (1977).
%%CITATION = PHRVA,D15,2738;%%
}

\lref\DS{
%\FabingerGP
M.~Fabinger and E.~Silverstein, ``D-Sitter space: Causal structure,
thermodynamics, and entropy,''
arXiv:hep-th/0304220.
%%CITATION = HEP-TH 0304220;%%
}

\lref\juan{
%\MaldacenaRE
J.~M.~Maldacena, ``The large N limit of superconformal field theories and
supergravity,'' Adv.\ Theor.\ Math.\
Phys.\ {\bf 2}, 231 (1998) [Int.\ J.\ Theor.\ Phys.\  {\bf 38}, 1113 (1999)]
[arXiv:hep-th/9711200].
%%CITATION = HEP-TH 9711200;%%
}

\lref\dSCFT{
%\StromingerPN
A.~Strominger, ``The dS/CFT correspondence,'' JHEP {\bf 0110}, 034 (2001)
[arXiv:hep-th/0106113]
%%CITATION = HEP-TH 0106113;%%
}

%\AcharyaKV
\lref\AcharyaKV{ B.~S.~Acharya, ``A moduli fixing mechanism in M theory,''
arXiv:hep-th/0212294.
%%CITATION = HEP-TH 0212294;%%
}

\lref\dSObj{
%\WittenKN
E.~Witten, ``Quantum gravity in de Sitter space,'' arXiv:hep-th/0106109;
%%CITATION = HEP-TH 0106109;%%
%\FischlerYJ
W.~Fischler, A.~Kashani-Poor, R.~McNees and S.~Paban, ``The acceleration of
the universe, a challenge for string
theory,'' JHEP {\bf 0107}, 003 (2001) [arXiv:hep-th/0104181];
%%CITATION = HEP-TH 0104181;%%
%\HellermanYI
S.~Hellerman, N.~Kaloper and L.~Susskind, ``String theory and
quintessence,'' JHEP {\bf 0106}, 003 (2001)
[arXiv:hep-th/0104180];
%%CITATION = HEP-TH 0104180;%%
%\GoheerVF
N.~Goheer, M.~Kleban and L.~Susskind, ``The trouble with de Sitter space,''
JHEP {\bf 0307}, 056 (2003)
[arXiv:hep-th/0212209].
%%CITATION = HEP-TH 0212209;%%
%\DysonNT
L.~Dyson, J.~Lindesay and L.~Susskind, ``Is there really a de Sitter/CFT
duality,'' JHEP {\bf 0208}, 045 (2002)
[arXiv:hep-th/0202163];
%%CITATION = HEP-TH 0202163;%%
%\BanksWR
T.~Banks, W.~Fischler and S.~Paban, ``Recurrent nightmares?: Measurement
theory in de Sitter space,'' JHEP {\bf
0212}, 062 (2002) [arXiv:hep-th/0210160];
%%CITATION = HEP-TH 0210160;%%
%\BanksYP
T.~Banks and W.~Fischler, ``M-theory observables for cosmological
space-times,'' arXiv:hep-th/0102077.
%%CITATION = HEP-TH 0102077;%%
}

\lref\KLT{
%\KrausHV
P.~Kraus, F.~Larsen and S.~P.~Trivedi, ``The Coulomb branch of gauge theory
from rotating branes,'' JHEP {\bf
9903}, 003 (1999) [arXiv:hep-th/9811120].
%%CITATION = HEP-TH 9811120;%%
}

%\AbbottQF
\lref\AbbottQF{ L.~F.~Abbott, ``A Mechanism For Reducing The Value Of The
Cosmological Constant,'' Phys.\ Lett.\
B {\bf 150}, 427 (1985).
%%CITATION = PHLTA,B150,427;%%
}

%\BanksMB
\lref\BanksMB{ T.~Banks, M.~Dine and N.~Seiberg, ``Irrational axions as a
solution of the strong CP problem in
an eternal universe,'' Phys.\ Lett.\ B {\bf 273}, 105 (1991)
[arXiv:hep-th/9109040].
%%CITATION = HEP-TH 9109040;%%
}

%\DasguptaSS
\lref\DasguptaSS{ K.~Dasgupta, G.~Rajesh and S.~Sethi, ``M theory,
orientifolds and G-flux,'' JHEP {\bf 9908},
023 (1999) [arXiv:hep-th/9908088].
%%CITATION = HEP-TH 9908088;%%
}

%\BrownKG
\lref\BrownKG{ J.~D.~Brown and C.~Teitelboim, ``Neutralization Of The
Cosmological Constant By Membrane
Creation,'' Nucl.\ Phys.\ B {\bf 297}, 787 (1988).
%%CITATION = NUPHA,B297,787;%%
}

%\BoussoXA
\lref\BP{ R.~Bousso and J.~Polchinski,
   ``Quantization of four-form fluxes and dynamical neutralization of the
cosmological constant,'' JHEP {\bf 0006}, 006 (2000) [arXiv:hep-th/0004134].
%%CITATION = HEP-TH 0004134;%%
}

%\ConlonDS
\lref\ConlonDS{ J.~P.~Conlon and F.~Quevedo, ``On the explicit construction and statistics of Calabi-Yau flux
vacua,'' JHEP {\bf 0410}, 039 (2004) [arXiv:hep-th/0409215].
%%CITATION = HEP-TH 0409215;%%
}

%\BalasubramanianUY
\lref\BalasubramanianUY{ V.~Balasubramanian and P.~Berglund, ``Stringy corrections to Kaehler potentials, SUSY
breaking, and the cosmological constant problem,'' arXiv:hep-th/0408054.
%%CITATION = HEP-TH 0408054;%%
}

%\GorlichQM
\lref\GorlichQM{ L.~Gorlich, S.~Kachru, P.~K.~Tripathy and S.~P.~Trivedi, ``Gaugino condensation and
nonperturbative superpotentials in flux compactifications,'' arXiv:hep-th/0407130.
%%CITATION = HEP-TH 0407130;%%
}

%\BlancoPilladoNS
\lref\BlancoPilladoNS{ J.~J.~Blanco-Pillado {\it et al.}, ``Racetrack inflation,'' arXiv:hep-th/0406230.
%%CITATION = HEP-TH 0406230;%%
}

%\FirouzjahiMX
\lref\FirouzjahiMX{ H.~Firouzjahi, S.~Sarangi and S.~H.~H.~Tye, ``Spontaneous creation of inflationary universes
and the cosmic landscape,'' JHEP {\bf 0409}, 060 (2004) [arXiv:hep-th/0406107].
%%CITATION = HEP-TH 0406107;%%
}

%\BuchbinderIM
\lref\BuchbinderIM{ E.~I.~Buchbinder, ``Raising anti de Sitter vacua to de Sitter vacua in heterotic M-theory,''
arXiv:hep-th/0406101.
%%CITATION = HEP-TH 0406101;%%
}

%\SchulzUB
\lref\SchulzUB{ M.~B.~Schulz, ``Superstring orientifolds with torsion: O5 orientifolds of torus fibrations and
their massless spectra,'' Fortsch.\ Phys.\  {\bf 52}, 963 (2004) [arXiv:hep-th/0406001].
%%CITATION = HEP-TH 0406001;%%
}

%\GiryavetsZR
%\lref\GiryavetsZR{ A.~Giryavets, S.~Kachru and P.~K.~Tripathy, ``On the taxonomy of flux vacua,'' JHEP {\bf
%0408}, 002 (2004) [arXiv:hep-th/0404243].
%%CITATION = HEP-TH 0404243;%%
%}

%\GiddingsVR
\lref\GiddingsVR{ S.~B.~Giddings and R.~C.~Myers, ``Spontaneous decompactification,'' Phys.\ Rev.\ D {\bf 70},
046005 (2004) [arXiv:hep-th/0404220].
%%CITATION = HEP-TH 0404220;%%
}

%\BergEK
\lref\BergEK{ M.~Berg, M.~Haack and B.~Kors, ``Loop corrections to volume moduli and inflation in string
theory,'' arXiv:hep-th/0404087.
%%CITATION = HEP-TH 0404087;%%
}

%\ChanMS
\lref\ChanMS{ C.~S.~Chan, P.~L.~Paul and H.~Verlinde, ``A note on warped string compactification,'' Nucl.\
Phys.\ B {\bf 581}, 156 (2000) [arXiv:hep-th/0003236].
%%CITATION = HEP-TH 0003236;%%
}

%\TownsendFX
\lref\TownsendFX{ P.~K.~Townsend and M.~N.~R.~Wohlfarth, ``Accelerating cosmologies from compactification,''
Phys.\ Rev.\ Lett.\  {\bf 91}, 061302 (2003) [arXiv:hep-th/0303097].
%%CITATION = HEP-TH 0303097;%%
}

%\BershadskyVM
\lref\BershadskyVM{ M.~Bershadsky, A.~Johansen, V.~Sadov and C.~Vafa, ``Topological reduction of 4-d SYM to 2-d
sigma models,'' Nucl.\ Phys.\ B {\bf 448}, 166 (1995) [arXiv:hep-th/9501096].
%%CITATION = HEP-TH 9501096;%%
}

%\DenefDM
\lref\DenefDM{ F.~Denef, M.~R.~Douglas and B.~Florea, ``Building a better racetrack,'' JHEP {\bf 0406}, 034
(2004) [arXiv:hep-th/0404257].
%%CITATION = HEP-TH 0404257;%%
}

%\FengIF
\lref\FengIF{ J.~L.~Feng, J.~March-Russell, S.~Sethi and F.~Wilczek,
``Saltatory relaxation of the cosmological
constant,'' Nucl.\ Phys.\ B {\bf 602}, 307 (2001) [arXiv:hep-th/0005276].
%%CITATION = HEP-TH 0005276;%%
}

%\KachruGS
\lref\KachruGS{ S.~Kachru, J.~Pearson and H.~Verlinde,
   ``Brane/flux annihilation and the string dual of a non-supersymmetric
field
%theory,''
JHEP {\bf 0206}, 021 (2002) [arXiv:hep-th/0112197].
%%CITATION = HEP-TH 0112197;%%
}

\lref\kls{kls refs}

\lref\gs{G. Dvali and S. Kachru, ``New Old Inflation,''
hep-th/0309095.}

\lref\SilversteinHF{E. Silverstein and D. Tong, ``Scalar Speed
Limits and Cosmology: Acceleration from D-cceleration,''
hep-th/0310221.}

\lref\jarv{L. Jarv, T. Mohaupt, and F. Saueressig, ``M-theory
cosmologies from singular Calabi-Yau compactifications,''
hep-th/0310174.}

\lref\bd{N. D. Birrell and P.C.W. Davies, {\it Quantum Fields in
Curved Space}, Cambridge University Press, Cambridge, England
(1982).}

\lref\horne{J. Horne and G. Moore, ``Chaotic Coupling Constants,''
   Nucl.Phys. {\bf{B}}432 (1994) 109, hep-th/9403058.}

\lref\hyb{A. Linde, ``Hybrid Inflation,'' Phys.Rev. {\bf{D}}49
(1994) 748, astro-ph/9307002.}

\lref\sw{N. Seiberg and E. Witten, ``Electric-Magnetic Duality,
Monopole Condensation, and Confinement in N=2 Supersymmetric
Yang-Mills Theory,'' Nucl.Phys. {\bf{B}}426 (1994) 19,
hep-th/9407087 \semi K. Intriligator and N. Seiberg, ``Lectures on
Supersymmetric Gauge Theories and Electric-Magnetic Duality,''
Nucl.Phys.Proc.Suppl. 45BC (1996) 1, hep-th/9509066.}

\lref\flux{some papers on moduli stabilization from fluxes}

%\draft

\Title{\vbox{\baselineskip12pt\hbox{hep-th/0411271} \hbox{SLAC-PUB-10859}\hbox{SU-ITP-4/41}}} {\vbox{
\centerline{A New Handle on de Sitter Compactifications}
\bigskip
\centerline{} }}

\centerline{Alex Saltman and Eva Silverstein \footnote{$^*$} {SLAC and
Department of Physics, Stanford
University, Stanford, CA 94309}}

\vskip .3in We construct a large new class of de Sitter (and anti de Sitter) vacua of critical string theory
from flux compactifications on products of Riemann surfaces.  In the construction, the leading effects
stabilizing the moduli are perturbative.  We show that these effects self-consistently dominate over standard
estimates for further $\alpha^\prime$ and quantum corrections, via tuning available from large flux and brane
quantum numbers.

\smallskip
%\draftmode
\Date{November 2004}

%\listtoc
%\writetoc

%\vfill \eject

\newsec{Introduction}

The construction and study of string theory models of de Sitter space is of interest for many reasons.   It
provides a basis for phenomenological models of dark energy and inflation from string theory and also provides a
concrete microphysical framework in which to investigate holography in the cosmological context.  The emerging
variety of discrete, physically connected solutions has interesting implications for the global structure of the
universe according to string theory and challenges conventional naturalness assumptions in the resulting
particle phenomenology.

In this paper, we present a new class of compactifications of
string theory, based on Riemann surfaces, yielding de Sitter (as
well as anti de Sitter) solutions in the resulting
four-dimensional effective field theory.  There are many potential
compactifications of string theory beyond those which classically
preserve a massless gravitino; those we consider here form a
particularly simple illustrative set of examples.  They realize
the case of Kaluza-Klein scale breaking of supergravity,
complementing the previous classes of models with sub-KK scale
supersymmetry breaking \kklt\ and string scale supersymmetry
breaking \mss.  This class turns out to be particularly simple,
involving basic aspects of the geometry of Riemann surfaces while
generating sufficiently generic contributions to the moduli
potential to meta-stabilize the system perturbatively.

Although generic vacua of this class will have high-scale
supersymmetry breaking, low-energy supersymmetric particle physics
models may be included, as the communication of the supersymmetry
breaking in the gravity sector allows for a separation of scales
between the matter superpartner masses and the KK-scale gravitino
masses.  In any case, the apparent proliferation of vacua in this
new class suggests that (as would be expected from genericity)
their number may significantly exceed that of the more symmetric
choices of vacua. We should also emphasize that this class is
itself likely to be a small corner of the space of possibilities.

Our construction starts from a compactification on a Riemann
surface of genus at least two whose complex-structure moduli we
stabilize via fluxes in a simple manner (explicitly for genus 2
and 3). Reducing from the critical dimension (for simplicity) this
leaves four remaining compact dimensions which can be compactified
in many ways. Perhaps the simplest option, which we exercise, is
to consider these to also be Riemann surfaces, giving us a
compactification on the product of three Riemann surfaces.

The low-energy theory obtained from a flux compactification on Riemann surfaces alone has remaining tadpoles for
the dilaton and the volume of each surface. In order to stabilize these moduli we add 7-branes, described
locally via an F-theory compactification on an elliptically fibered fourfold. We argue that, by a generalization
of a similar mechanism used in \refs{\gkp,\kklt}, the set of intersecting wrapped 7-branes (equivalently the
topology of the F-theory fourfold) provides an anomalous negative 3-brane charge and tension and therefore a
tunable negative contribution to the potential.  More specifically, in order to obtain a net negative
contribution to this part of the moduli potential and the three-form fluxes, while satisfying the constraints of
Gauss' law in the compactification, we need to includes sets of 7-branes separated from sets of anti 7-branes,
which have a similar anomalous negative 3-brane tension (without net charge), and behave in our stabilized
regime as a generalization of combinations of O3 planes and anti O3-planes. This procedure retains the general
property of KK scale breaking of supergravity, while providing needed forces to stabilize the volumes and
dilaton perturbatively.

We will at various points make genericity arguments, for example regarding the positions that the 7-branes
settle and regarding a technical simplification of the axion contribution to the dynamics.  These arguments are
based on contributions such as ambient fluxes generic to our construction but we will not exhibit their effects
in detail (for the 7-brane moduli, our treatment is thus similar to \refs{\gkp,\kklt}). In any case, we will
provide an explicit perturbative stabilization mechanism for the dilaton and volume moduli, as well as
Riemann-surface complex-structure moduli.

\subsec{Relations to previous works}

The previous string-theoretic models of de Sitter fall into two classes--those based on supersymmetric
low-energy effective theories arising from Calabi-Yau compactifications of critical string theory \kklt, and
those based on string-scale supersymmetry breaking in supercritical limits of string theory \mss.\foot{Many
important works have appeared recently in the area of moduli stabilization and the discretuum, including for
example \refs{\BP,\ChanMS, \achar,\splitsusy,\BeckerGW,\us,\distributions, \GVW,
\SusskindKW,\renata,\KST,\FreyHF,\TripathyQW,\DasguptaSS,\TaylorII,\MayrHH,\ferrara,\MarchesanoYN, \ConlonDS,
\BalasubramanianUY, \CamaraJJ, \BlumenhagenVR,\GorlichQM, \BlancoPilladoNS}
\refs{\FirouzjahiMX,\BuchbinderIM,\SchulzUB, \GiddingsVR, \BergEK, \susyscale, \gia}.  The works for example
\refs{\TownsendFX,\ohta}\ considered compactification on hyperbolic spaces, and the work \BershadskyVM\
considered compactifications of field theories on Riemann surfaces.} The present models lie in between these
classes. Here supersymmetry is generically broken in the gravity sector at an intermediate scale corresponding
to the Kaluza-Klein scale of the compactification. This may still allow for low-energy supersymmetry in the
matter sector, but with intermediate scale gravitini\foot{This may be of use in addressing the gravitino
problem, as discussed recently in \luty\AR\ and references therein.}; more generally these models have
non-supersymmetric spectra at low energies (and some of them may fit into the framework \splitsusy). The
complex-structure moduli of the Riemann surface are stabilized by one-form fluxes on pairs of dual one-cycles in
a way similar to the stabilization of complex-structure moduli of the Calabi-Yau in \gkp\ by three-form fluxes
on pairs of dual three-cycles.  The dilaton tadpole present at leading order in the expansion in string coupling
is of the same form as that arising in supercritical limits of perturbative string theory \mss. Like the models
\mss, Riemann-surface compactifications exhibit sufficient forces to stabilize all moduli perturbatively, while
like the models \kklt, the species enhancement to the effective couplings is manifestly controllable.

\subsec{A note on control.}

For readers most familiar with low-energy-supersymmetric compactifications, it is worth reviewing the methods
for theoretical control that exist in the absence of supersymmetry below the Kaluza-Klein scale.  We use a
controlled perturbation expansion in the low-energy effective field theory derived from string theory.  This is
obtained by introducing by hand sufficiently large flux and brane quantum numbers to ensure that couplings are
stabilized at small enough values and volumes at large enough values that the solution has small flux energy
density and small effective couplings (including enhancements from numbers of species running in loops). This
procedure, and the control it affords, does not depend on low-energy supersymmetry.\foot{Indeed, low-energy
supersymmetric non-renormalization theorems, if too powerful, can preclude the generation of sufficient forces
to fix moduli perturbatively. Thus they can require the tuning of classical effects against non-perturbative
effects, a procedure that turns out to be possible (and elegant) \kklt\ but is arguably harder than the tuning
needed to play different orders of perturbation theory off of each other. Conversely, with {\it either} ${\cal
N}=1$ supersymmetry {\it or} ${\cal N}=0$, the moduli potential suffers from quantum corrections at arbitrary
loop orders, and hence perturbative control must be established in much the same way in both cases.}

As in flux compactifications on spheres (familiar in recent years
for their role in the AdS/CFT correspondence), the tadpoles
arising from the curvature in the geometry are cancelled by forces
introduced by other ingredients arising at higher orders in the
expansion in inverse volumes and string coupling (such as fluxes,
wrapped branes and orientifolds). It is not necessary for
consistency to cancel tadpoles order by order in string
perturbation theory--indeed such a procedure would guarantee that
the dilaton is {\it not} fixed perturbatively.

\newsec{Riemann surface flux compactification}

Consider string theory compactified on a Riemann surface $\Sigma$
of genus $h$.  The light degrees of freedom (which we will refer
to as ``moduli") arising from the metric on the Riemann surface
are as follows. Using diffeomorphism invariance, we can reduce the
metric degrees of freedom to the complex structure moduli, which
are zero modes, the conformal factor, whose overall volume mode we
will keep and whose higher KK modes we will self-consistently
ignore, and off-diagonal metric modes, which we will ignore
because they are massive due to the absence of continuous
isometries in higher-genus Riemann surfaces.

As an example, one can consider a configuration in which the
 metric has constant curvature on the Riemann surface--this is the configuration toward which the system
evolves in the absence of other sources, or for sources which are
uniformly distributed on the surface. For $h\ge 2$ the surface has
$3h-3$ complex-structure moduli which correspond to changes of
metric that do not affect the curvature, and hence are massless
deformations about that configuration. In addition, there is a
scalar field corresponding to the volume of the surface, arising
from the lowest mode of the conformal factor of the metric.

We will introduce fluxes and other ingredients to stabilize these moduli as well as the other potentially
runaway moduli (such as the dilaton) arising in a full compactification down to four dimensions.  In particular,
we will obtain solutions with a local minimum of the potential energy above zero, i.e. metastable de Sitter
solutions. Because the Kaluza-Klein modes start with large masses from their internal gradients, we will ignore
their dynamics here; in our final analysis we will see that the KK scale can be tuned to be parametrically
higher than the curvature scale of the four-dimensional spacetime, and that the masses for the moduli can be
arranged to be parameterically lighter than the KK masses in the minimum if so desired.

The Einstein term $\sqrt{g} {\cal R}$ integrated over the Riemann
surface produces a tree-level contribution to the low-energy
effective potential proportional to $2h-2$.  In four-dimensional
Einstein frame (appropriate after further compactification on a
space $X$ of volume $V_X$ in string units) this contribution
scales like\foot{See \modlectures\ for a basic review of the
potential energy arising from various ingredients in string
compactifications.}
\eqn\curvpotI{U_{\cal R}\sim {1\over l_4^4} (2h-2) {g_s^2\over {V_\Sigma^2 V_X}}}
where $l_4$ is the four-dimensional Planck length. The contribution \curvpotI\ provides a tree-level force on a
combination of the dilaton and volume moduli, but does not depend on the $3h-3$ (for $h\ge 2$) complex moduli of
the Riemann surface.

More generally, we will consider 7-branes embedded in $\Sigma$, described locally via F-theory on an appropriate
fourfold geometry. These contribute positive potential energy from their tension in addition to negative
contributions arising from F-theoretic curvature couplings which we will review later.

In standard F-theory constructions, one considers F-theory compactified on an elliptically-fibered Calabi-Yau
manifold, which in the corresponding type IIB description is a set of intersecting 7-branes embedded in the base
of the fibration \fpapers.  The positive tension contributions of the 7-branes add to the tree level tadpole in
\curvpotI. For example, the type IIB description of F-theory on K3 amounts to 24 (p,q) 7-branes embedded in a
genus $h=0$ compactification, in such a way that the 7-branes cancel the term \curvpotI\ completely. More
generally, if we view a higher-genus Riemann surface $\Sigma$ as a $\IP^1$ with handles attached, we can for
example consider the same set of 7-branes on $\Sigma$. In the F-theory description, we are patching into the
base of the fibration a trivial fibration of a $T^2$ over a set of handles.  Allowing for multiple such sets of
24 (including separated antibrane sets in general), this means that \curvpotI\ becomes
\eqn\curvpot{U_{{\cal R},7Bs}\sim {1\over l_4^4} (2h+2(n_7-1)) {g_s^2\over {V_\Sigma^2 V_X}}}
where $n_7$ represents the number of sets of 24 7-branes (and/or anti 7-branes) included in the surface. Here we
used that the $h=0$ contribution in \curvpotI\ is cancelled by the effects of one such set of sevenbranes; this
is a good approximation at large volume, where the SUSY breaking scale is much smaller than the 7-brane
tensions. More generally, we can consider additional D7-branes and anti-D7-branes; their effects on the
potential energy will arise at the next order in $g_s$. In general, we will find the need for sets of both
7-branes and anti 7-branes; we will discuss their position moduli in \S3.

Fluxes threading one-cycles of $\Sigma$ will prove useful for stabilizing the volume and dilaton, and also yield
classical forces on the complex-structure moduli from the flux kinetic terms ${\cal L}_{flux}\sim -\int F\wedge
\ast F$; we will study this explicitly in the next subsection. The basic physics is as follows: flux through a
one-cycle forces the cycle to expand to lower the energy density contained in the flux. Similarly, flux through
the dual to this cycle forces the dual cycle to expand. At fixed total volume, the combination of these two flux
effects tends to stabilize the ratio of the sizes of the cycles and their duals. As we will see in \S2.1, in
order to achieve this stabilization for all the independent complex structure moduli, we will require $2h$
independent fluxes threading the one-cycles of $\Sigma$.

\subsec{Calculating the flux potential}

We would like to compute the potential energy obtained from the
flux kinetic energy for one-form fields threading one-cycles on a
Riemann surface. We do this explicitly for the cases of genus
$h=1,2,3$, for which the period matrix (reviewed below) provides a
faithful representation of the Riemann-surface complex-structure
moduli space. This enables us to organize the problem as a simple
change of basis between the integral basis appropriate to
quantized fluxes and the holomorphic basis defining the period
matrix.  Our analysis may generalize to higher-genus examples if
the extra directions in the period-matrix description can be dealt
with.

Consider a genus $h$ Riemann surface $\Sigma$ with homology basis
$a_i, b_i$ and integral one-forms $\alpha_i,\beta_i$ such that
\eqn\duals{\eqalign{
    \int_{a_i} \alpha_j &= \delta_{ij} \cr
    \int_{b_i} \beta_j &= \delta_{ij} \cr
    }
}
and for any one-form $\eta$
\eqn\integrals{\eqalign{
    \int_{a_i} \eta &= \int_\Sigma \eta \wedge \beta_i \cr
    \int_{b_i} \eta &= \int_\Sigma \alpha_i \wedge \eta. \cr
    }
}
We wish to calculate the potential $U_{\rm flux} = \int_\Sigma F
\wedge \ast F$ for a one-form field $F$ in terms of the vector
$(m_i,n_i)$ such that $F = m_i \alpha_i + n_i \beta_i$ (up to a
shift by an exact 1-form). This becomes cleaner by a
transformation to a holomorphic basis where the Hodge star is
easily defined. We can take a standard basis of $h$ holomorphic
one-forms $\omega_i$ with antiholomorphic partners
$\overline{\omega}_i$ such that
\eqn\periods{\eqalign{
    \int_{a_i} \omega_j &= \delta_{ij} \cr
    \int_{b_i} \omega_j &= \tau_{ij} \cr
    }
}
for some symmetric period matrix $\tau$ with ${\rm Im} \, \tau$
positive definite.

Considering our flux one-form in the $\omega$ basis and in the $\alpha,\beta$ basis
\eqn\changebasis{
    F = u_i \omega_i + \overline{u}_i \overline{\omega}_i = m_i \alpha_i + n_i \beta_i+{\rm exact},
}
we can see that
\eqn\transform{\eqalign{
    \int_{a_i} F &= \delta_{ij} u_j + \delta_{ij} \overline{u}_j = m_i \cr
    \int_{b_i} F &= \tau_{ij} u_j + \tb_{ij} \overline{u}_j = n_i, \cr
    }
}
so we can define a $2h \times 2h$ matrix $K$ such that
\eqn\Kdef{\eqalign{
    K &= \pmatrix{I & I \cr \tau & \tb} \cr
    \pmatrix{m \cr n} &= K \pmatrix{u \cr \overline{u}}. \cr
    }
}

Now the Hodge star acts on forms in the $\omega$ basis as
\eqn\hodge{\eqalign{
    \ast \omega &= -i \omega \cr
    \ast \overline{\omega} &= i \overline{\omega} \cr
    }
}
so it transforms the coefficients through the action of the matrix
$H$ defined as
\eqn\Hdef{
    H = \pmatrix{-i I & 0 \cr 0 & i I}.
}
Also, from \integrals\ we know that $\int_\Sigma \alpha_i \wedge \beta_j = \delta_{ij}$, so altogether we get
\eqn\potential{
    U_{\rm flux} = (m \,\, n) M K H K^{-1} \pmatrix{m \cr n}
}
where
\eqn\Mdef{
    M = \pmatrix{0 & I \cr -I & 0}.
}

To calculate everything in terms of $\tau$, we need to invert $K$. We find
\eqn\Kinverse{
    K^{-1} = \pmatrix{-(\tau-\bar\tau)^{-1}\bar\tau & (\tau-\bar\tau)^{-1} \cr (\tau-\bar\tau)^{-1}\tau &
    -(\tau-\bar\tau)^{-1}}
}
Multiplying, we get
\eqn\multiply{\eqalign{
    M K H K^{-1} &=
         i \pmatrix{ 2 \tau (\tau-\tb)^{-1} \tb & -(\tau + \tb)(\tau-\tb)^{-1} \cr
                -(\tau-\tb)^{-1} (\tau + \tb) & 2 (\tau-\tb)^{-1}}. \cr
    }}
For $h=1$ this reduces to
\eqn\torus{
    M K H K^{-1} = {2 i\over{\tau - \tb}} \pmatrix{|\tau|^2 & -{\rm Re} \, \tau \cr -{\rm Re} \, \tau &
    1}.
}

So we obtain a flux potential energy
\eqn\fluxpot{U_{\rm flux}\propto \sum_{I=1}^{N_F} Q^{iI} A_i^j(\tau) Q_j^I }
where $I$ indexes $N_F$ different types of flux with one component
on $\Sigma$, $i$ and $j$ index the quantum numbers on the $a$ and
$b$ cycles, and $A_i^j(\tau)$ is the $2h\times 2h$ matrix
\eqn\Amatrix{A(\tau) =  i \pmatrix{ 2 \tau (\tau-\tb)^{-1} \tb &
    -(\tau + \tb)(\tau-\tb)^{-1} \cr
                -(\tau-\tb)^{-1} (\tau + \tb) & 2 (\tau-\tb)^{-1}}}
In \fluxpot\ we have indicated the full dependence on $\tau$,
while further dependence on volumes will arise in a complete
construction via further compactification (and conversion to
four-dimensional Einstein frame) in a way to be discussed
explicitly below.

If we have higher-form fluxes that wrap cycles on more than one
Riemann surface, the flux quantum numbers will just gain extra
indices to be contracted by $A$, e.g. for a three form flux
wrapping one-cycles of three different Riemann surfaces, the
potential contribution is
\eqn\fluxpotgen{
    U_{\rm flux} \propto Q^{ikm} A_i^j(\tau_1) A_k^l(\tau_2) A_m^n(\tau_3) Q_{jln}
}
where $\tau_1, \tau_2, \tau_3$ are the period matrices of the three surfaces and the flux indices can vary over
different ranges if the genera of the Riemann surfaces are different.

\subsec{Analysis of $\tau$ stabilization}

We would like to understand if the potential energy \fluxpot\ is
sufficient to metastabilize the complex-structure moduli $\tau$.
If we can show that the potential blows up at all the boundaries
of the moduli space, then it must have a minimum in the
interior.\foot{In fact, since the one-form fluxes spontaneously
break modular invariance, at fixed flux quantum numbers we are
interested in the boundary of the covering space of the
Riemann-surface moduli space.} These boundaries are known--they
correspond to the imaginary part of $\tau$ becoming degenerate or
$\tau$ factoring into smaller Riemann surfaces (e.g. $\tau_{12}
\rightarrow 0$ in the $h=2$ case.) The first boundary can be dealt
with fairly easily if there are $2h$ independent flux vectors:
Since $A$ is a positive-definite symmetric matrix, it has all
positive eigenvalues. As an eigenvalue of $\tau-\tb$ approaches 0,
$(\tau-\tb)^{-1}$ has an eigenvalue that gets large. This
dominates the other components of $A$, meaning that $A$ must get a
large eigenvalue, but with $2h$ independent fluxes at least one
has a component along the eigenvector with large eigenvalue, and
thus the inner product $\fluxpot$ goes to infinity.

This argument does not deal with the factorization boundary, which is of a different character.  The first type
of boundary involves the shrinking of a nontrivial cycle, which some flux wraps, while the factorization limit
involves the shrinking of a trivial cycle, which no flux wraps. However, for the case of small genus, we can
explicitly find local minima of the flux potential for $\tau$ away from all boundaries. At $h=3$, a computer
search of 100 linearly-independent flux choices found minima away from the boundary in all cases. It is possible
to vary the flux choices, which allows some degree of tuning of the solutions for $\tau$. The computer search
demonstrated a wide variation in the locations, which suggests that they are indeed tunable. It would be nice to
have a conceptual argument for why the factorization boundary is avoided--a similar effect occurs in the case of
the large complex structure limit of flux-stabilized Calabi-Yau compactifications \gkp.

\newsec{The volume and dilaton tadpoles: basic strategy}

Having fixed the complex-structure moduli via the flux potential described in the last section, we now turn to
the stabilization of the dilaton and volume moduli.  As discussed above, we have a tree-level tadpole for these
quantities after compactifying on $\Sigma$.  For our case of $h\ge 2$, this tree-level contribution to the
potential energy is positive, driving the volume toward large values and the dilaton toward weak coupling.  We
need to introduce further ingredients in order to stabilize these directions, while also stabilizing additional
runaway moduli introduced by the new ingredients, and doing it all in a way consistent with a controlled
perturbative approximation scheme. This (in our experience) requires some trial and error, which results in the
class of de Sitter models which we present in the next section. However, much of the input is based on simple
intuition about the forces at play in compactifications with flux and branes \modlectures, so we will start in
this section by sharing our basic strategy.

Because all sources of Einstein-frame potential energy go to zero at weak coupling and large volume,
stabilization requires sufficiently strong negative contributions.  To obtain de Sitter space we aim for three
terms in increasing orders of perturbation theory about weak coupling and large volume, with the middle term
negative.  For example, for the string coupling $g$, fixing the other moduli at their ultimate minima we require
a potential of the form $ag^2-bg^3+cg^4$ which for large enough $b$ and $c$ produces a metastable minimum at
weak coupling. In our examples, the tree-level contribution \curvpot, and \fluxpot\ in the case of NS 3-form
flux, produces a positive term proportional to $g^2$. We can add a negative term at order $g^3$ (the order at
which orientifolds and anti-orientifolds, the simplest such negative contribution, arise) and use the RR flux
appearing at order $g^4$ to provide a final positive term.

In fact, we will now argue that there is a more general way to obtain a {\it tunably large} negative
contribution generalizing that discussed in \S2.1\ of \gkp, by using the fact that wrapped intersecting branes
and related curvature contributions can produce negative D3-brane tension and charge.

Consider first the low-energy supergravity case of \gkp, starting
from F-theory compactifications on elliptically fibered Calabi-Yau
fourfolds. The base of the fibration has 7-branes at the loci
where the $T^2$ fiber degenerates. There is an anomalous
contribution to the D3-brane charge, given for a Calabi-Yau
fourfold by $-\chi_4/24$ where $\chi_4$ is the Euler character of
the fourfold. There is correspondingly a supersymmetric partner of
this charge, an effective 3-brane tension also given by
$-\chi_4/24$.  By varying the choice of fourfold, one can tune
this contribution to large negative values. (In the IIB language
this tunably-large anomalous charge and corresponding tension is
associated in large part to inflow on intersections of 7-branes as
in \inflow.) For perhaps the simplest case of a CY-fourfold
fibered over $\IP^1\times \IP^1\times\IP^1$, $\chi_4/24=732$
\KlemmTS, which yields a control parameter of order $10^{-3}$;
more general fibrations can yield $\chi_4/24$ significantly larger
\kreuzer.

In the Calabi-Yau case there is a positivity condition among a
subset of contributions to the potential.\foot{We thank R.
Blumenhagen and F. Denef for reminding us of this important
relation and the resulting need for the more generic ingredients
to follow.} Specifically the NSNS and RR 3-form fluxes $H_3$ and
$F_3$ are related to the anomalous 3-brane charge by Gauss's law:
\eqn\gausslaw{
    {1\over (2 \pi)^4 (\alpha')^2} \int_{\Sigma^3}H_{(3)} \wedge F_{(3)} = Q_{3,7}
}
where $Q_{3,7}$ is the anomalous 3-brane charge of the 7-branes. In the low energy supersymmetric case, the
anomalous charge is related by a BPS condition to the anomalous negative tension.  This implies \gkp\ that the
combination of the $H_3$ and $F_3$ flux potential terms and the anomalous negative term is positive
semidefinite.

But in the non-supersymmetric case there will generically be both 7-branes and anti-7-branes. Adding extra sets
of 7-branes and anti-7-branes to the base leads to triple intersections which locally each preserve 1/8 of the
ambient supersymmetry, and hence behave identically to the local contributions in a CY-fourfold; i.e. they
contribute to the effective 3-brane tension. However, these pairs of branes will carry no net 3-brane charge,
which will excuse us from the positivity relation that applies in the SUSY case\foot{They will also break
supersymmetry at the KK scale, just like the Riemann surfaces themselves.}. Specifically, if we add extra sets
of 7-branes and anti 7-branes, each in the same combination of 24 appearing in F-theory on K3, then the net
negative contribution to the tension follows from adding these local sources and there will be no contribution
to \gausslaw. More generally, we can also consider the possibility of extra $D7-\overline{D7}$ contributions,
which will contribute to the same localized three-brane tension term.

Adding branes leads, a priori, to further brane moduli in need of stabilization, as in \refs{\gkp,\kklt}.
Although we add both branes and antibranes, there is a way to include them which minimizes their mutual
interactions. Namely, in both the supersymmetric and non-supersymmetric cases, appropriate combinations of 6
(p,q) 7-branes do not carry net 7-brane charge (the combinations corresponding to $SO(8)$ points on K3).  So
such sets of 6 branes separated from sets of 6 antibranes avoid extra gauge attraction.  As a result, these
defects behave, from the point of view of each surface, as a set of massive sources in the effective 2+1
dimensional gravity system. These effective masses get contributions from both the 7-brane and anti 7-brane
tensions (included for $n_7$ (p,q) sets of 24 in \curvpot), and the anomalous negative tension we just reviewed
which appears at the next order in $g_s$ perturbation theory. Note that because of the anomalous negative
contribution, there is a large negative offset to their effective masses in the 2+1 gravity problem on each
surface. In particular, in the presence of both branes and antibranes, one might worry that they will tend to
annihilate. However, with the negative contribution in place, the annihilation will not be energetically
preferred in expansion about our solution. This is reminiscent of the fact that orientifolds and
antiorientifolds do not annihilate (indeed a more prosaic, but less tunable, contribution at the same order
would come from explicit $O3-\overline{O3}$ pairs).

So we will introduce such sets of branes and antibranes to obtain
a tunably large negative tension contribution independent of the
charge. But as in the original low-energy supersymmetric models
\refs{\gkp, \kklt}\ we will not explicitly stabilize the 7-brane
moduli, leaving this to the genericity of the flux contributions
and other corrections which can locally lift these non-chiral
modes.\foot{See  \DDFK\ for an explicit model in the low energy
SUSY context eliminating the 7-brane moduli.}  Though we strongly
expect generic contributions to indeed lift these non-chiral
moduli, this will be the least explicit part of our construction.

So far we have only discussed Calabi-Yau F-theory models, with addional localized sources. Let us now return to
our main interest, type IIB string theory in the critical dimension, compactified on a product of Riemann
surfaces. Let us consider, for example, the above case of a CY fourfold fibered over $\IP^1\times
\IP^1\times\IP^1$, but sew in handles to the base $\IP^1$ factors in smooth regions away from the 7-branes, as
well as adding extra sets of 7-branes and anti-7-branes as just discussed. For a general (non-Calabi-Yau)
fourfold, the anomalous 3-brane charge is proportional to \refs{\svw,\beckersX}
\eqn\fullcharge{-{1\over 768}(tr (R^2))^2 + {1\over 192}tr R^4}
In the regions containing the handles, this contribution vanishes
because the manifold is a product of the handle times a 3-manifold
in that region and the traces in \fullcharge\ vanish for this
configuration (to be specific, one may consider adding handles and
brane-antibrane pairs to the CY base in its orientifold limit
\refs{\senF,\GopakumarMU,\keshavetal}). As a result, the tunably
large negative contribution to the 3-brane tension survives to a
good approximation. It is only approximate because supersymmetry
is broken, and the local relation between the anomalous 3-brane
charge and tension is corrected. However, at large volume the
supersymmetry is broken at a scale much lower than the scale of
the 3-brane tension, so the BPS relation between the anomalous
contribution to the charge and that of the effective 3-brane
tension is still a good approximation as long as we ultimately
stabilize the system in the large-volume regime.

Altogether, these ingredients yield a term in the effective potential of the form
\eqn\threecont{U_3\sim -{1\over l_4^4}N_7{g_s^3\over V^2} }
where $V$ is the total volume of the compactification, and $N_7$
is the effective control parameter just described coming from the
triple intersections of 7-branes. In particular $N_7$ has a
contribution scaling like $(24 n_7)^3/24$ in the notation of
\curvpot.  More generally, there are other contributions. One
could add $n_{D7}$ sets of $D7-\overline{D7}$s, which would lead
to contributions to $N_7$ scaling like cubic combinations of
$n_{D7}$ and $n_7$. One could also include explicit
$O3-\overline{O3}$ pairs.  In writing down \threecont\ we have
used the fact that the anomalous contributions do not depend on
any geometric moduli in the string frame (and only depend on the
overall volume in via the volume factor coming from the conversion
to four dimensional Einstein frame).

In what follows we will include 7-branes via a term of the form
\threecont, including also their positive 7-brane tension as in
\curvpot, and show how the volumes and dilaton are stabilized. We
will not here explicitly determine where the 7-branes become
localized inside the compactification, we expect ambient fluxes,
dS thermal effects, and quantum corrections to generically lift
these moduli. As discussed above, by combining the branes and
antibranes each in appropriate groups of $(p,q)$ objects we can
avoid strong forces between them.

\newsec{A new class of de Sitter models}

With the above inputs and motivations, we present our class of de Sitter models.

\subsec{The ingredients}

Consider type IIB string theory in the critical dimension compactified on a product of three Riemann surfaces
$\Sigma_s$ of genera $h_s$, $s=1,2,3$.  On each Riemann surface, we insert 7-branes and anti-7-branes yielding a
net positive contribution to $N_7$, as defined above, with enough 7-brane-anti-7-brane pairs to obtain a tunably
large net negative contribution to the part of the potential coming from the 3-form fluxes to be included and
\threecont. Each 7-brane sits at a point in one of the $\Sigma_s$ factors and wraps a four cycle consisting of
the other two Riemann surfaces. The control parameter $N_7$ behaves cubically in the 7-brane number, a feature
that will figure into our analysis of the size of corrections. For simplicity, we will often take $h_1 = h_2 =
h_3 = h$ in the following.

In addition, we include the following types of flux in the
compactification.  First some notation:  let $i_s=1,\dots 2h_s$
index the 1-cycles on $\Sigma_s$.  As above, let $s=1,2,3$ index
the Riemann surface factors $\Sigma_s$.  The ingredients are:

\medskip

\noindent (1)  Neveu-Schwarz 3-form flux $H_3$ on 3-cycles consisting of one one-cycle in each of the three
$\Sigma$ factors.  Let us denote these flux quantum numbers $N^{i_1 i_2 i_3} $.

\noindent (2)  Ramond-Ramond 3-form flux $F_3$ on 3-cycles consisting of one one-cycle in each of the three
$\Sigma$ factors.  Let us denote these flux quantum numbers $Q_3^{i_1 i_2 i_3} $.

\smallskip

The contributions (1) and (2) must satisfy the constraint that $\int H_3\wedge F_3$ cancel the tadpole in
three-brane charge arising from the net intersecting wrapped 7-branes described above \gausslaw.

\smallskip

\noindent (3)  Ramond-Ramond 1-form flux on the 1-cycles of
$\Sigma_s$; the corresponding flux quantum numbers will be denoted
$Q_1^s$.  This contribution will not play a significant role in
our stabilization mechanism, but can be included among this class
of models.

\noindent (4)  Ramond-Ramond 5-form flux on the 5-cycles
consisting of 1-cycles of $\Sigma_s$ times $\Sigma_{s+1} \times
\Sigma_{s+2}$ where the subscripts are reduced mod 3.  The
corresponding flux quantum numbers will be denoted $Q_5^{I_s}$.

There is an interesting subtlety with this contribution.  As discussed in \fluxorientifold, in an orientifold
limit of the system of 24 7-branes on $\IP^1$, $T^2/(I_2\Omega (-1)^{F_L})$, we cannot consider 5-form flux with
one index transverse to the O7-plane.  That is, the zero mode of this flux is projected out by the orientifold
action. More precisely, the flux must vanish at the positions of the O7-planes; all the KK excitations of this
sort (with zeros at the $O7$-planes) are projected in and are hence consistent configurations.

Of course Kaluza-Klein excitations of the RR flux would not be suitable for our model building, since such modes
necessarily fluctuate in time. Fortunately in our case we add handles to the $\IP^1$, and the topology of the
Riemann surfaces come to the rescue.  The holomorphic 1-forms, which correspond to static solutions of the
equations of motion for the flux, have zeros on higher genus Riemann surfaces (at genus $h$ the holomorphic
1-forms each have $2h-2$ zeros by the Riemann-Roch theorem). Starting from a configuration where we sew handles
onto the orientifold limit of the $\IP^1$ with 24 7-branes, in the $\Sigma_i$ direction in which our 5-form flux
reduces to a 1-form we must consider the 1-form with zeros at the positions of the orientifold planes.  The flux
solution is determined by the integer flux quanta and the complex structure only up to cohomology, so the zeros
can be placed at the correct points by adding exact 1-forms. Having addressed this constraint in the orientifold
limit, the setup should be consistent for more general configurations:  nonsingular deformations of the
compactification away from this limit will not change topological features such as the number of zeros.  More
generally, fluxes with boundary conditions that they must vanish at the positions of certain defects may be
accommodated by considering sufficient genus to obtain enough zeroes in the corresponding holomorphic 1-forms.
In general cases, this requires our flux stabilization in \S2\ to persist for higher genus.

\medskip

\subsec{The low-energy spectrum}

In what follows we will analyze the potential for the potentially
runaway moduli coming from the dilaton and the the volumes of the
three Riemann surfaces.  Because of the absence of isometries, we
do not expect massless off diagonal components in the metric, both
scalars that would modify the product structure of our base
manifold and massless gauge fields in four dimensions arising from
the $10d$ metric. However there are other light degrees of freedom
in the low-energy effective field theory arising from our
construction. In particular, there are RR gauge fields and axions
from dimensional reduction of the 10 dimensional RR potentials on
the cycles of our compactification.

Let us consider the axions. These come from the type IIB RR scalar $C_0$ and harmonic 2-forms made from the NS
and RR potentials $B_2$ and $C_2$ integrated over 2-cycles of the compactification, as well as the RR 4-form
$C_4$ on 4-cycles. The IIB axion $C_0$ couples via the term
\eqn\Czerocoup{{U}_{C_0} \propto \int |F_3-C_0 H_3|^2}
In general, one can solve $C_0$'s equation of motion and plug the result into \Czerocoup, yielding a more
complicated potential energy for the other moduli (including $\tau$) than that arising in the absence of the
$C_0$ coupling. This potential may generically have nontrivial local minima, but we can simplify the situation
by arranging the fluxes to produce a solution at or near $C_0=0$. That is, we can arrange that the $C_0$ tadpole
cancel by setting to zero the coefficient of its linear term in \Czerocoup
\eqn\Czerotad{{U}_{C_0 ~ tadpole}\propto C_0\int (H_3\wedge\ast F_3 +
    F_3\wedge\ast H_3)\equiv 0.}
This condition is consistent with our requirement to simultaneously turn on $\int H_3\wedge F_3$ to satisfy the
Gauss' law constraint \gausslaw.

The axions arising from the 2-forms $B_2$ and $C_2$ have similar couplings:
\eqn\twocouplings{{U}_{C_2,B_2}\propto \int |F_5-{1\over 2}C_2\wedge H_3+{1\over 2}B_2\wedge F_3|^2}
We can also arrange a solution with negligible tadpoles for $C_2$
and $B_2$ by insisting that the coefficient of the linear term for
these axions ($\int_{2-cycles} C_2$ and $\int_{2-cycles} B_2$) be
small.  The number of conditions on the flux choices this entails
is $2b_2$ where $b_2=12h^2+3$ (given the same genus $h$ for all
three Riemann surfaces).  The number of 3-form fluxes alone is $16
h^3$ so it is possible to accommodate these conditions, leaving
behind enough independent fluxes to have $2h$ independent fluxes
threading one-cycles on each Riemann surface factor. Similar
comments apply to the $C_4$ field, whose contribution scales like
that of $C_2$.

If we made more general choices than those yielding $C_0\sim 0\sim
\int B_2\sim \int C_2\sim\int C_4$, we would obtain a somewhat
more complicated effective potential for the other moduli.  This
may be interesting to study. In any case, having made these
simplifications in the pseudoscalar sector, we move on to the
potential energy for the volume moduli and the dilaton.

\subsec{The potential energy}

We combine the sources described in the previous sections,
including \curvpot\fluxpot\threecont. We will denote the volume in
string units of the $\Sigma_s$ factors $V_s$.
\eqn\fullpot{\eqalign{
    U = &{1\over l_4^4} {g_s^4\over (V_1 V_2 V_3)^2}\biggl\{
    \sum_{r} {1\over g_s^2} 2 (h_r+n_7-1)V_{r+1} V_{r+2} +
    {1\over g_s^2}N^{i_1 i_2 i_3} A(\tau_1)_{i_1}^{j_1} A(\tau_2)_{i_2}^{j_2} A(\tau_3)_{i_3}^{j_3} N_{j_1 j_2 j_3} \cr
    & - {1\over g_s} N_7
    + Q_3^{i_1i_2i_3} A(\tau_1)_{i_1}^{j_1}A(\tau_2)_{i_2}^{j_2}A(\tau_3)_{i_3}^{j_3} Q_{3~j_1j_2j_3} \cr
    & + \sum_{r} \biggl[ Q_1^{i_r} A(\tau_r)_{i_r j_r} Q_1^{j_r}  V_{r+1} V_{r+2}
    + Q_5^{i_r} A(\tau_r)_{i_rj_r} Q_5^{j_r} {1\over {V_{r+1} V_{r+2}}} \biggl] \biggr\} \cr }
}
where we take the $r$ index to range cyclically over the labels
1,2,3 of the three Riemann surface factors. Here the overall
factor comes from the conversion to four-dimensional Einstein
frame and the first term is from the Einstein-Hilbert action,
dimensionally reduced as in \curvpot. We have included the offset
from the 7-branes adding to the $2h-2$ coefficient of \curvpotI.
The second term contains the NS flux contribution, as in
\fluxpotgen; the next is the effective 3-brane tension
contribution from intersecting wrapped 7-branes, as in \threecont.
The last terms give the RR flux contributions. We have dropped
several order-one factors; we will keep track of the dependence on
our discrete parameters which will provide parametric control.

\subsec{Metastable minima of the moduli potential}

In this subsection, we will elucidate how this model, with appropriately tuned choices of flux and brane quantum
numbers, produces metastable de Sitter minima.  We will first recall the result of \S1\ that (with sufficient
independent fluxes) the complex moduli of each $\Sigma$ are metastabilized. Then we will observe that e.g. for
an approximately symmetric distribution of fluxes, the relative volumes $V_r/V_s$ get stabilized at order 1 by
the potential \fullpot.  Finally, we will demonstrate the stabilization of the overall volume and dilaton.

These manipulations will lead us to tuning requirements on the
flux and brane quantum numbers, as well as expressions for how the
stabilized values of the moduli scale with these discrete quantum
numbers. In the following subsection we will use these results to
make standard estimates for the size of the $\alpha^\prime$ and
$g_s$ corrections to the background, which must be tuned small
self-consistently.  We will see that the required tuning is
available in our system given the possibility of scaling up the
contribution $N_7$ as discussed above.

\noindent{\it Complex moduli of $\Sigma$}

In \S1, we established that the complex-structure moduli $\tau$ of the Riemann surface are stabilized by $2h$
independent fluxes threading one-cycles of $\Sigma$.  The 1-form and 5-form fluxes alone are sufficient to
stabilize the complex moduli of a genus 2 Riemann surface. More generally, the 3-form fluxes lead to $4 h_{r+1}
h_{r+2}$ types of fluxes on $\Sigma_r$, allowing one to stabilize the complex structure moduli of products of
higher genus surfaces as well.

\noindent{\it Ratios of $\Sigma$ volumes}

In \fullpot, the positive terms in the potential are symmetric among the three $\Sigma$ factors.  Let us scale
out the dependence on the overall volume $V = V_1V_2V_3$ in each term.  The positive terms in the resulting
potential all go to infinity for any large ratio of volumes $V_r/V_s$.  If we tune the fluxes to be
approximately symmetric among the three $\Sigma_r$ factors, then we obtain a minimum at $V_r/V_s \sim 1$ from
these terms\foot{More generally, one can use the symmetry of the potential and the Arithmetic Mean-Geometric
Mean inequality to show that the volumes are stabilized at values of order the ratios of flux numbers.}.

\noindent{\it Overall volume and dilaton}

Let us now move to the problem of stabilizing the overall volume $V=V_1V_2V_3$ and dilaton.  Setting the
relative volumes $V_r/V_s\equiv 1$ our potential reduces to one of the form
\eqn\Uvoldil{\eqalign{ U(g_s,V) & = C\biggl(g_s^2 (h+n_7-1) {1\over V^{4/3}} + g_s^2 n_3^2{1\over V^2} \cr &
-g_s^3 N_7{1\over V^2} \cr & + g_s^4 q_3^2{1\over V^2} + g_s^4 q_1^2 {1\over V^{4/3}}+ g_s^4 q_5^2{1\over
V^{8/3}} \biggr) \cr }}
Here for simplicity we have taken the fluxes to be symmetrically distributed among the three $\Sigma$ factors,
each of genus $h$. We have introduced the shorthand $n_3^2$ for the flux potential from the NS 3-form, evaluated
at the minimum $\tau=\tau_*$, and similarly for the Ramond fluxes $q_a^2$.
%
%\eqn\fluxanswers{\eqalign{
%& N^{Ia}A(\tau)_{IJ} N^{Ja}|_{min}\equiv n_a^2, ~~~ a=1,2 \cr & Q_1^I A(\tau)_{IJ}
%Q_1^J|_{min}\equiv q_1^2 ~~~ Q_5^I A(\tau)_{IJ}Q_5^J|_{min}\equiv q_5^2 }}
%

Consider first the first, third, and sixth terms in \Uvoldil.
These form a fourth degree polynomial in the combination
$g_s/V^{2/3}$ so for appropriate choices of coefficients, they
stabilize this combination at
\eqn\firstset{{g_s\over V^{2/3}} \sim {h+n_7-1\over N_7}\sim
{N_7\over q_5^2}}
Plugging this into the remaining terms (the second, fourth, and fifth) in \Uvoldil, we obtain a contribution
\eqn\lastset{U_{2-4-5}\sim \biggl({h+n_7-1\over N_7}\biggr)^2 n_3^2{1\over V^{2/3}}+\biggl({h+n_7-1\over
N_7}\biggr)^4q_3^2 V^{2/3}+ \biggl({h+n_7-1\over N_7}\biggr)^4q_1^2 V^{4/3}}
If we consider flux choices such that the first two terms here
dominate (while remaining subdominant to the first, third and
sixth terms discussed above), the second set of terms will not
destabilize $g_s / V^{2/3}$ and we find $V$ stabilized such that
\eqn\Vresult{ V^{4/3} \sim \biggl({{N_7n_3}\over {(h+n_7-1)q_3}}\biggr)^2}
while the string coupling is stabilized at
\eqn\gresult{g_s\sim {n_3\over q_3}}

This procedure for stabilizing the volume and dilaton requires
that the terms \lastset\ are subdominant to the first, third and
sixth terms in \Uvoldil.  In particular, the 3-form NS and RR flux
potentials at the minimum, $n_3^2$ and $q_3^2$, need to be
sufficiently small so that the corresponding terms are
subdominant. As discussed above, at the same time, we must satisfy
\gausslaw, which ties the flux quantum numbers $N_3$ and $Q_3$ to
net 7-brane contributions to the effective D3-charge.  Given the
combinations of branes and antibranes in our construction, which
contribute to the anomalous negative tension but not the charge,
this is not a problem.

\noindent{\it The cosmological constant and moduli masses}

The set of metastable minima we have exhibited produces a discretuum of possible cosmological constants,
depending on the flux choices.  If we do not tune coefficients significantly to obtain a low cosmological
constant, then the scale of the resulting cosmological constant is of order the Kaluza-Klein scale of the
compactification
\eqn\ccscale{\Lambda_{untuned}\sim {h\over V^{1/3}}{1\over\alpha^\prime}}
By tuning our discrete parameters we can arrange the cosmological constant and moduli masses to produce a
hierarchy of scales between the KK scale of the compactification and the curvature scale in four dimensions as
well as the moduli mass scales:
\eqn\tuninggoals{\Lambda_{tuned}, m^2_{moduli}\ll \Lambda_{untuned}}

\noindent{\it Anti de Sitter examples}

Although we have focused on de Sitter minima, similar methods lead
to anti-de Sitter vacua.  For example, in situations where the
negative 7-brane contribution is sufficiently strong relative to
the RR flux contributions, our minimum may dip below zero.

To obtain de Sitter solutions, a necessary condition is at least
three independent contributions in each direction in an expansion
about weak coupling/large volume, with the middle term negative.
More generally, to obtain parts of the discretuum that are purely
anti-de Sitter, we can consider other compactifications which only
provide two terms (negative, positive) in some or all the
directions.  This allows one to consider products of Riemann
surfaces with spheres or orbifolds of spheres.

\subsec{Estimates of subleading corrections and self-consistency checks}

In the above analysis, we exhibited metastable minima with small
string coupling and large volumes (relative to the string scale).
Although this is a necessary condition for a controlled solution,
there are further self-consistency checks we must make.  In the
presence of large flux and brane quantum numbers, it is necessary
to check that the effective expansion parameters coming into
stringy and quantum corrections are small.  These expansion
parameters are somewhat enhanced by the large discrete quantum
numbers.  Hence in this subsection, we will systematically
estimate the size of the corrections taking these factors into
account.

\noindent{\it Curvature corrections}

String theory has a generic expansion in $\alpha^\prime {\cal R}$
where ${\cal R}$ is the curvature of the spacetime background. By
tuning \Vresult\ large, we can preclude these large curvature
corrections in our type IIB background. Specifically, in string
units,
\eqn\curvcorr{
    {\cal R} \sim {h \over V^{1/3}} \sim {h(h+n_7-1)^{1/2} \over
    N_7^{1/2}} \biggl({q_3 \over n_3}\biggr)^{1/2} \ll 1.
}
Keeping in mind that $q_3 / n_3 \sim 1/g_s$, this requires that $N_7$ be large.

Note that as in \refs{\gkp,\kklt}, we are using a large number of 7-branes to obtain a strong negative
contribution in the potential energy. In F-theory (or related M-theory or IIA backgrounds) this is itself
related to an ${\cal R}^4$ correction. It would be nice to check explicitly whether there are any other enhanced
quartic curvature contributions, and if so if these are independently tunable.  This question also arises in the
low energy supersymmetric models, and we believe it is reasonable to take the approach of \gkp\kklt, and assume
that these other ${\cal R}^4$ contributions are subdominant since the large number introduced by the
intersecting branes contributing to $\chi_4$ need not generally contribute coherently to other curvature terms.
Indeed, this statement was checked for some situations in \beckers. Note, as discussed above, that adding
handles to the base in itself adds a small correction to the $\chi_4$ ${\cal R}^4$ contribution.

\noindent{\it NS flux corrections}

Next let us consider the expansion in $ (H_3 \alpha^{\prime})^2 \sim n_3^2/V$.  To make this small we require
\eqn\Hcondition{\biggl({(h+n_7-1) \over N_7}\biggr)^{3/2} n_3^{1/2} q_3^{3/2} \ll 1}
which is weaker than \curvcorr\ since we have already insisted that $n_3 q_3 \ll N_7$.

\noindent{\it RR flux corrections}

The RR flux vertex operators come with an additional factor of
$g_s$.  Hence the condition for control of the higher derivative
terms involving a $p$-form flux is $g_s^2 q_p^2/V^{p/3} \ll 1$.
For the threeform flux $F_3$, this is satisfied in a similar way
to that described above for $H_3$ \Hcondition.

For the 5-form RR flux we obtain the condition
\eqn\fivecond{
    {N_7 g_s \over V} \sim {(h+n_7-1)^{3/2} \over N_7^{1/2}} \biggl({q_3 \over n_3}\biggr)^{1/2} \ll 1
}
which is the strongest condition we encounter.  Since $N_7\sim 24^2 n_7^3$, this contribution is of order
$g_s^{-1/2} 10^{-3/2}$. This permits reasonable control if we consider also a $g_s$ of order 1/10.  If one
includes additional contributions coming from $D7-\overline{D7}$ pairs in such a way that they contribute
positively to $N_7$, this control becomes parameteric.

\noindent{\it Quantum corrections}

Finally, we need to consider the expected strength of quantum corrections to our Lagrangian in this background.
At high energy-momentum flowing through the loops, above the KK scale, the contributions are localized in the
compactification and the effective coupling is the 10d coupling $g_*\sim n_3/q_3$ \gresult. We therefore choose
$q_3$ at least somewhat larger than $n_3$.  At low energies, in the 4d effective field theory, the corrections
scale like $g^2/V$ times the number of species running in loops. This number is roughly $N_7$ at low energies
(if we keep the genus $h$ low enough that topological enhancements from the handles are subdominant). Putting
this together with the above scalings, we have
\eqn\quantumcorr{{N_7 g_s^2\over V} \sim {(h+n_7-1)^{3/2} \over N_7^{1/2}} \biggl({n_3 \over q_3}\biggr)^{1/2}
\ll 1.}
This is weaker than \fivecond\ by a factor of $g_s$.

\newsec{Discussion}

This class of models provides a perturbative set of de Sitter vacua to a good approximation, as long as the
non-chiral 7-brane moduli are lifted as expected by the ambient fluxes and quantum corrections (an issue common
to this case and the low energy supersymmetric models). In particular, we exhibited sufficient controlled forces
to stabilize the string coupling and volumes, as well as the Riemann surface complex structure moduli,
perturbatively. It further illustrates the fact that the string theory ``landscape" goes beyond the low energy
supergravity sectors most studied to date (though these models may still permit low-energy supersymmetry in the
matter sector).)  In particular, in studying the discretuum of string vacua, it does not suffice to consider
only those with a low-energy supergravity effective Lagrangian in four dimensions.

\subsec{Number of Vacua}

As anticipated in \BP, like other examples of flux vacua our construction leads to the possibility of mass
production of metastable string vacua, by varying over the many possible flux, brane, and topological quantum
numbers. Let us estimate roughly the number of vacua in our new class of models. As discussed in \BP, a rough
estimate for the number of vacua is obtained by counting the volume in a sphere in flux space up to a maximum
radius determined by the strength of the ``bare" negative cosmological constant to be cancelled by the fluxes.
In our case, as in \kklt\ the ``bare" negative piece is dominated by the 7-brane contribution in \fullpot. This
is cancelled in part by flux squared contributions, so we can roughly trade this for $M^2 $ where $M$ scales
like the dominant flux quantum numbers. Let us denote the maximal value of this quantity (determined by the
maximal value of $N_7$ available including back reaction constraints) by $M_{max}$. Our space of fluxes is
$2b_3+b_1+b_5$ dimensional (where $b_p$ are the numbers of noncontractible $p$-cycles in the compactification).
Putting this together, our estimate for the number of vacua available here is
\eqn\nvac{N_{vac}\sim \biggl({M^2_{max}\over {b_3+(b_1+b_5)/2}}\biggr)^{b_3+(b_1+b_5)/2}}
Note that relative to the Calabi-Yau case, the extra handles enhance the dimension of the flux space.  In
general, one might expect that relaxing conditions such as the Calabi-Yau condition enhances the number of
independent ingredients, hence increasing the number of vacua in the more generic starting points.  In \nvac\ we
see one aspect of that here.  Of course the examples we have studied themselves constitute only a small corner
of the space of compactifications.

It would be interesting to determine the distributions of these
vacua in moduli space, as in \distributions. It is clear that the
assumption of low energy supersymmetric effective Lagrangians must
be relaxed in order to obtain a representative sample of string
vacua, and to answer related questions about the statistics of the
supersymmetry-breaking scale \refs{\BP,\mss,\susyscale}. This
point, already clear from the case of string scale supersymmetry
breaking, is reinforced by the intermediate class of models
discussed here.

\subsec{Holographic Duality}

The new de Sitter construction may also be of interest for studying the microphysics of de Sitter space, since
the perturbative ingredients involved in the moduli stabilization, including the volume stabilization, are
fairly explicit. As before, the fluxes can be traded for branes to expose some of the microphysical content of
the holographic duals on their approximate moduli space \coulombduals.  In this regard it is interesting that as
in the construction \kklt, here we required nontrivial NS flux $H_3$, which leads to the presence of NS-branes
in the approximate moduli space of the dual.  It would be interesting to know if this is part of a general
pattern.

\subsec{String Duality}

These compactifications raise interesting questions concerning
string-string duality. One possible way to explore this would be
to elucidate more explicitly whether there is a useful F theoretic
description of $T^2$ fibrations over more generic base manifolds
(such as the $\Sigma^3$ in the present construction).

Another possibility is to try to understand the small-radius behavior of compactifications on Riemann surfaces.
In the present work, we tuned to obtain a set of large-radius compactifications in order to maintain general
relativity as a good approximation. However, the small-radius limit of the Riemann surface may remain a
well-defined conformal field theory if we take into account strong worldsheet dynamics (and remain on shell by
including the time dependence arising from the tadpoles, or by including other ingredients to metastabilize the
space). It is even possible that the theory grows dimensions at strong worldsheet coupling; this possibility is
perhaps suggested by the fact that the form of the tree-level dilaton tadpole in our theory \curvpotI\
\eqn\treeterm{U_{\cal R}\sim {1\over l_4^4}(2h-2) {g_s^2\over {V_\Sigma^2 V_X}}}
is reminiscent of that in dimension $D$ supercritical string theory models \dealwisetal\mss
\eqn\treetermsupercrit{U_{D}\sim {1\over l_4^4}(D-10){g_s^2\over V_{internal}}}
In both these formulas there is an integer quantum number in the coefficient--which serves as the effective
central charge which must be soaked up by dilaton time dependence or higher order balancing of forces. It is
related to the first Chern class of the manifold in the case of Riemann-surface compactifications, and is
related to the dimensionality in the case of supercritical limits of string theory.  It would be very
interesting to understand if these two integers are related to each other by a duality, for example in the small
radius regime of the Riemann surface compactifications.

\subsec{Phenomenological applications}

Finally, it would be interesting to explore the phenomenology of these models.  One question to ask is whether
low-energy SUSY in the particle physics sector may emerge in some models of the sort we consider here (cf. for
example \luty). In particular, gravitational communication of the SUSY breaking of the gravity sector to the
Standard Model leads to superpartner mass squares of order $m_{KK}^4/M_p^2$. This alone would provide $TeV$
scale SUSY breaking in the observable sector if $m_{KK}\sim 10^{11} ~GeV$, though there may be other mediation
mechanisms which dominate this depending on the details. In any case, more generically, we expect many models
with high-scale SUSY breaking in this context; it would be interesting to check for models of the sort
\splitsusy. String-theoretic standard-model constructions based on intersecting branes on tori might fit into
this framework well, as locally there are products of circles in our $\Sigma^3$ compactifications.\foot{We thank
M. Schulz for this comment.}

\medskip

\noindent{\bf Acknowledgements}

We thank P. Aspinwall, R. Blumenhagen, F. Denef, S. Dimopoulos, M. Dine, L. Fidkowski, J. Hsu, A. Kashani-Poor,
X. Liu, L. McAllister, J. McGreevy, A. Pierce, M. Schulz, C. Vafa, and especially S. Kachru and A. Tomasiello
for many useful discussions. We thank the organizers and participants of the String Vacuum Workshop in Munich,
and the SLAC Theory group, for useful comments when this work was first presented.  We are supported in part by
the DOE under contract DE-AC03-76SF00515 and by the NSF under contract 9870115.

\listrefs

\end